\documentclass[nofootinbib, aps, prd, a4paper, 10pt, superscriptaddress, eqsecnum, showkeys]{revtex4-2}
\usepackage[a4paper, top=2cm, bottom=2cm, left=2.5cm, right=2.5cm]{geometry}

\usepackage{amsmath}
\usepackage{amsfonts}
\usepackage{amsthm}
\usepackage{bm}
\usepackage{mathrsfs}
\usepackage{braket}

\usepackage{graphicx}
\usepackage{booktabs}

\usepackage{xcolor}
\usepackage[pdfusetitle]{hyperref}
\hypersetup{colorlinks = true, allcolors = blue}

\usepackage{orcidlink}

\begin{document}

\title{Is MOND an Emergent Effective Law of a Complex Dark Matter Sector?}
\author{V.K.~Oikonomou\orcidlink{0000-0003-0125-4160}}
\email{voikonomou@gapps.auth.gr} \affiliation{Department of
Physics, Aristotle University of Thessaloniki, Thessaloniki 54124,
Greece} \affiliation{Center for Theoretical Physics, Khazar
University, 41 Mehseti Str., Baku, AZ-1096, Azerbaijan}
\author{Eleni I. Manouri\orcidlink{0009-0007-8243-6340}}
\email{elenimanouri21@gmail.com;emanouri@physics.auth.gr}
\affiliation{Department of Physics, Aristotle University of
Thessaloniki, Thessaloniki 54124, Greece}

\begin{abstract}
In this work we aim to present a new perspective for MOND theory,
namely that MOND theory is not an answer for the missing matter at
galactic scales, but it provides a phenomenological description of
the dynamics that DM should produce at galactic scales. We also
aim to point out that MOND dynamics at galactic scales is achieved
by scale-dependent self-interacting dark matter, which can behave
as collisionless and collisional, depending on the scales and the
physical processes. To this end, we first aim to highlight the
successes of the $\Lambda$-Cold-Dark-Matter model and how it
overwhelms over the MOND paradigm. Using textbook physics, we
present the successes of the $\Lambda$-Cold-Dark-Matter at
cosmological and cluster scales, and how these phenomena cannot be
described consistently by MOND theories. Thus, by excluding MOND
as being a viable description of nature at all scales, we conclude
that MOND may not be the answer behind missing matter in Newtonian
galactic dynamics, but it is just showing how the dynamics of dark
matter should behave at galactic scales.
\end{abstract}

\maketitle

\section{Introduction}

MOND (Modified Newtonian Dynamics) theory \cite{milgrom1983} is a
theoretical framework, which proposes the modification of Newton's
law of gravity, and at galactic scales results to the successful
explanation of the galactic rotation curves and the baryonic
Tully-Fisher relation. At galactic scales it is a successful
theory and eliminates the need for cold dark matter (CDM).
Nowadays MOND theory is considered as an alternative to CDM and
some consider it an alternative to the $\Lambda$-Cold-Dark-Matter
($\Lambda$CDM) model. However, MOND theory is successful only at
galactic scales, while at cluster and super-cluster scales and at
cosmological scales, MOND utterly fails to provide a successful
phenomenological description of nature. In most cases, the MOND
supporters rely on extra matter content, apart from the baryonic,
to explain the Bullet cluster physics or even cosmological
evolution using neutrinos. Eventually MOND was engineered to solve
dynamics without DM, and with the inclusions of extra matter
components, the MOND theory uses the same conceptual approach as
the $\Lambda$CDM, it uses missing matter to explain phenomena.
Thus from a conceptual point of view, it has no advantage over the
$\Lambda$CDM, apart from being particularly simple compared to the
$\Lambda$CDM, and it is way less technical and more accessible to
a broader scientific audience.

In this work we aim to review the phenomenological successes of
the $\Lambda$CDM model, which MOND cannot explain at all. We
additionally aim to consider MOND not as an alternative theory to
CDM but consider it as some phenomenological characteristic of DM.
In fact, CDM has shortcomings at galactic scales, for example the
cusp-core problem, the too-big-to-fail problem and also the
diversity problem \cite{Tulin:2017ara}. Thus CDM is challenged at
galactic scales, where MOND theory solves the problems with hands
down. On the other hand, self-interacting DM usually solves many
of the CDM problems \cite{Tulin:2017ara}. Thus this makes us think
that MOND theory is not a theory itself, but it might be a
characteristic of DM at galactic scales. Hence, DM at galactic
scales could have the dynamical characteristics of MOND theory,
while at cluster, supercluster and cosmological scales, DM behaves
as CDM. In this line of reasoning it is evident that DM might have
a scale dependent EoS, behaving as self interacting at galactic
scales
\cite{Slepian:2011ev,Saxton:2014swa,Alonso-Alvarez:2024gdz,Kaplinghat:2015aga,Ahn:2004xt,Saxton:2012ja,Saxton:2010jk,Saxton:2016ozz,Arbey:2003sj,
Heikinheimo:2015kra,Wandelt:2000ad,Spergel:1999mh,Loeb:2010gj,Ackerman:2008kmp,Goodman:2000tg,Arbey:2006it,Yang:2025ume,Abedin:2025dis},
while at larger scales it might behave as CDM. This sort of
behavior would require a scale-dependent equation of state (EoS)
as described and analyzed in Refs.
\cite{Oikonomou:2026vkp,Oikonomou:2025bsi}. This sort of behavior
can easily be modelled by mirror DM
\cite{Kobzarev:1966qya,Hodges:1993yb,Foot:2004pa,Berezhiani:2003wj,Silagadze:2008fa,Foot:2000tp,Chacko:2005pe,Berezhiani:2000gw,Blinnikov:2009nn,Mohapatra:2001sx,
Blinnikov:1982eh,Blinnikov:1983gh,Foot:2016wvj,Foot:2014osa,Foot:2014uba,
Foot:2004pq,Foot:2001ft,Foot:2004dh,Foot:1999hm,Foot:2001pv,Foot:2001ne,Foot:2000iu,Pavsic:1974rq,Foot:1993yp,Ignatiev:2000yy,Ignatiev:2003js,
Ciarcelluti:2004ik,Ciarcelluti:2004ip,Ciarcelluti:2010zz,Dvali:2009fw,Foot:2013msa,Foot:2013vna,Cui:2011wk,Foot:2015mqa,Foot:2014mia,Cline:2013zca,Ibe:2019ena,
Foot:2018qpw,Howe:2021neq,Cyr-Racine:2021oal,Armstrong:2023cis,Ritter:2024sqv,Mohapatra:1996yy,Mohapatra:2000qx,Goldman:2013qla,Berezhiani:1995am,Oikonomou:2024geq},
which can be comprised by elementary particles, atoms and heavier
elements. Thus mirror DM can actually mimic self-interacting DM
behavior at galactic scales, while at larger scales it can mimic
CDM. Hence our proposal is simple, MOND is not an autonomous
theory that can describe galactic scale phenomena, but it
describes the phenomenological behavior of DM at galactic scales.
Hence, with this work we aim to show explicitly, using textbook
physics, why MOND cannot be considered an autonomous theory of
gravity, and we shall try to motivate the conceptual proposal that
MOND is actually describing phenomenological laws of DM at
galactic scales. Thus MOND is not an alternative to DM, it
describes its dynamics at galactic scales only.

Let us recall in brief the conceptual framework of MOND theory,
which is particularly simple. At low accelerations gravity behaves
differently than what Newton and Einstein described. Essentially
in MOND theories, Newton's laws breaks down at accelerations below
$a_0\approx1.2\times10^{-10}$ m/$s^2$, in order to explain the
flat galactic rotation curves, which contradict the Keplerian
decline expected from the visible baryonic mass distribution
itself. The original formulation of MOND indicates that the true
acceleration $a$ which is experienced by a test particle is
related to the Newtonian acceleration $a_N$, by an interpolating
function $\mu$:
\begin{equation}
    a \cdot\mu \left(\frac{a}{a_0}\right)=a_N\, ,
\end{equation}
with $a$ being the true acceleration, $a_N$ being the Newtonian
acceleration expected from visible baryonic mass, $a_0$ is
Milgrom's constant and $\mu(x)$ is the interpolating function,
which must have the following asymptotic behavior
\cite{milgrom1983},
\begin{itemize}
    \item When $x\gg 1$, $\mu(x) \rightarrow 1$ (Newtonian physics)
    \item When $x \ll 1$, $\mu(x) \rightarrow x$ (leads to $a=\sqrt{a_Na_0}$)
\end{itemize}
The $\Lambda$CDM on the other hand, is a solid paradigm, which
explains our observations regarding the evolution of the Universe
by invoking invisible forms of matter, the CDM, and invisible
forms of energy, the dark energy quantified by the cosmological
constant $\Lambda$. Its components, the cosmological constant and
the CDM part explain a plethora of phenomena, including the flat
galactic rotation curves, the patterns in the cosmic microwave
background (CMB), the accelerated expansion of the Universe and so
on. The $\Lambda$CDM assumes that the Universe's total
matter/energy content is divided into three components,
\begin{itemize}
    \item $5\%$ Baryonic (ordinary) matter: The visible matter that all celestial objects are made off.
    \item $27\%$ CDM (cold DM): The matter which explains the galactic rotation curves, the gravitational lensing,
    and the large-scale structure formation.
    \item $68\%$ Dark energy quantified by the cosmological constant $\Lambda$. This is the generator which drives the accelerated expansion of the Universe.
\end{itemize}
These percentages have been measured and constrained by CMB
experiments like the Planck collaboration \cite{Planck:2018vyg},
and the $\Lambda$CDM model successfully reproduces the
observations of the CMB, the large-scale galactic distribution,
and the abundances of the light elements from the Big Bang
Nucleosynthesis (BBN).

Thus with this work we aim to review and highlight the important
scientific successes of the $\Lambda$CDM, which MOND cannot
address successfully, with the most important ones being the
baryon acoustic oscillations, the CMB anisotropies and structure
formation, through the cosmological perturbation theory. These
physical phenomena present a strong challenge for MOND theories.
Also we shall discuss the hierarchical evolution of structure
formation and that the Milgrom's constant is thus redshift
dependent, which contradicts MOND. Combining all the evidence we
present here, we shall conclude with the fact that the
$\Lambda$CDM model is a superior paradigm
\cite{Planck:2018vyg,Planck:2019nip,BOSS:2016wmc}. However, we aim
to generate an alternative way of thinking regarding MOND theory,
namely that MOND is not an answer to galactic scale phenomena, but
it is a phenomenological characteristic of DM at galactic scales.
We shall devote a section regarding this perspective.

Finally, let us note that MOND theory lacks of a consistent
relativistic description that is compatible with the GW170817
event. There exist relativistic descriptions like the
tensor-vector theories \cite{Bekenstein:2004ne}, however these are
incompatible with GW170817. Recently however, several relativistic
formulations of MOND theory were developed and these are
interesting to note
\cite{Deffayet:2024ciu,Boran:2017rdn,Deffayet:2014lba,Deffayet:2011sk,Deffayet:2025lwl}.
In these works a concrete proof of the fact that MOND-mimicking
modified gravity can reproduce the successes of the $\Lambda$CDM
model was not firmly given. In \cite{Deffayet:2025lwl}  the
reconciliation of the non-local theories with $\Lambda$CDM model
is achieved, but the approach is not detailed enough and remains
at description level, without examining thoroughly the CMB peaks
and behavior of the theory. We also need to note that the
statements which concern the limitation of MOND theories, which
are developed in the following sections, refer mainly to a pure
modified gravity setting, without assuming additional matter
components. So pure MOND is considered, not MOND with missing mass
added other than DM. Thus, our results do not indicate that every
relativistic MOND realization fails, we just discuss that MOND
without missing matter fails at super-galactic and cosmological
scales, and at the same time we show that a successful theory that
realizes relativistically the MOND paradigm must literally pass
through the eye of the needle to prove viable phenomenologically.
Therefore, whenever we refer to MOND theories hereafter we refer
to pure modified gravity MOND theories, without additional matter
components, since MOND with additional matter is not a strict MOND
theory, it is similar in spirit to the $\Lambda$CDM.

In view of our scale-dependent DM, these relativistic formulations
of MOND might be a means of describing scale-dependent EoS DM
phenomena using modified gravity. Modified gravity is known to
mimic DM at several scales
\cite{Jusufi:2023ayv,Capozziello:2012ie,Lubini:2011pc,Capozziello:2008oya,Martin-Moruno:2008qpc,Capozziello:2006uv,Capozziello:2005tf,Capozziello:2006ph,Capozziello:2020dvd,Bernal:2011qz,Bernal:2011wm,Capozziello:2017rvz}.

The paper is organized as follows: In section II we focus on the
BAO and the need for particle non-baryonic DM, in section III we
discuss the CMB Anisotropies, the Peaks of CMB and the need for
DM. In section IV we analyze the CDM and structure formation while
in section V we attempt a direct comparison of MOND Theories and
the $\Lambda$CDM Model, and at section VI we present direct
evidence of DM coming from dark galaxies and notable galactic
clusters. In section VII we discuss the spin problem of galaxies
and how DM resolves it, while in section VIII we aim to present a
new viewpoint of MOND theories, namely that MOND is not an actual
theory that describes nature at galactic scales, but it is merely
a description of particle DM at these scales, and also we show
that a self-interacting DM with a scale-dependent EoS can be
compatible with the MOND postulates at galactic scales.

The discussions and presentation of the following sections aims to
establish several observational requirements that any viable DM
scenario must satisfy in order for it to be considered viable.
That said, we do not aim to gradually construct independent
evidence for the scale-dependent EoS perspective which is
developed in section VII, which stands alone. Our approach is
clear, we aim to present all the successes of particle DM over the
MOND paradigm, in order to show that MOND is merely successful at
galactic scales only, and then we introduce our central idea, that
MOND is not a theory by itself that can describe nature at
galactic scales, but it is a description of what should DM obey at
galactic scales. This leaves room for self-interacting DM with a
scale-dependent EoS, which is our proposal. Specifically, the
cosmological and the cluster-scale phenomena we shall present in
the following sections, require a gravitating non-baryonic
particle component, which must be that of collisionless DM at the
scales examined. The aim thus of sections II-VI is merely to
establish the foundational requirements for DM which are imposed
by the CMB, the BAO, the structure formation, the cluster dynamics
and the gravitational lensing. This approach aims to support the
need of DM, its undoubtable presence through the dynamics, and to
pave the stage for our new proposal, that MOND is not a
self-standing theory but could be viewed as a property of DM at
the galactic scales. Overall thus, the proposal we aim to
introduce in this work, does not replace essentially the
cosmological role of CDM by MOND and thus our proposal is a
question whether the successful MOND phenomenology at galactic
scales could arise as an effective property of particle DM which
is self-interacting and scale-dependent at galactic scales.


\section{Baryon Acoustic Oscillations and the Need for DM}

We will start reviewing the successes of the $\Lambda$CDM with the
baryon acoustic oscillations, see for example the textbooks
\cite{lyth2009,padmanabhan2002}. Baryon acoustic oscillations
(BAOs) and their physics primary confirms the need for CDM.
Essentially BAOs are the frozen imprint of sound waves from the
early Universe, which can still be observed in the large scale
distribution of galaxies today. They constitute the measure of the
vast distances in the cosmos, while through them we are able to
further understand the history of its expansion. The story starts
at the recombination era, where the Universe is dense and opaque,
a plasma soup made of tightly coupled baryons, electrons and
photons with the baryons being tightly coupled to photons by
Thomson scattering. Within this primordial plasma, tiny ripples
existed, namely the cosmological perturbations generated by the
inflationary era. At this point, the cosmic fluid consists of four
components, baryons, CDM, photons and neutrinos and also we assume
that CDM and neutrinos have negligible interaction. Let us
consider the photon decoupling epoch in which there are free
electrons and nuclei, while also photons are scattered off
electrons via direct Compton scattering. At this point, the
tightly coupled baryon-photon fluid, upon horizon entry,
oscillates as a standing acoustic wave. This is negligible on
small scales which are defined by the Silk scale, which represents
the characteristic length below which the acoustic oscillations
are damped by the photon diffusion, an effect which erases
perturbations at small scales. Before the recombination era, even
though that baryons and photons are tightly coupled via Thomson
scattering, photons perform random walks through the
baryon-electron fluid. Essentially the Silk scale is $k_S^{-1}$
and it is the distance a photon can diffuse up until
recombination. Qualitatively, perturbations smaller than the Silk
scale are suppressed exponentially, and this is why we obtain a
damping tail in the CMB power spectrum at high multipoles.
Considering scales larger than the Silk scale, the anisotropies of
the CMB give us acoustic oscillations on the last scattering
sphere. Also at the photon decoupling, baryons are free to fall
into the CDM's density perturbation gravitational potential wells.
This is very important and we further discuss it later on. The
standing acoustic wave which is created, is essentially the baryon
density perturbation and in order to work with the baryon pressure
we obtain the equations for the separate evolution of the CDM
(which has negligible pressure) and the baryonic matter (we are in
the Newtonian gravity regime so this makes sense only after
decoupling) \cite{lyth2009}. The Euler and Poisson equations for
CDM are respectively \cite{lyth2009}:
\begin{equation}
    \dot\delta_c + \frac{k}{a}V_c = 0\, ,
\end{equation}
\begin{equation}
    \dot V_c +HV_c -\frac{k}{a}\Phi = 0\, ,
\end{equation}
while the Euler and Poisson equations for baryonic matter are
respectively,
\begin{equation}
    \dot\delta_B + \frac{k}{a}V_B = 0\, ,
\end{equation}
\begin{equation}
    \dot V_B +HV_B - \frac{k}{a}\Phi = \frac{k}{a} \frac{\delta
    P_B}{\rho_B}\, .
\end{equation}
Using the above, the evolution equation for the baryonic density
contrast is obtained,
\begin{equation}
    \ddot\delta_B + 2H\dot\delta_B -\frac{3}{2}H^2\delta + \left(\frac{k}{a}\right)^2c_s^2\delta_B =
    0\, ,
\end{equation}
where $c_s^2\equiv\delta P_B / \delta\rho_B$ and $\delta$ is the
total matter density contrast, basically the effect of gravity on
the evolution of $\delta_B$. Now for the CDM we obtain
equivalently:
\begin{equation}
    \ddot\delta_c + 2H\dot\delta_c - \frac{3}{2}H^2\delta = 0\, .
\end{equation}
Here $\delta$ is the effect of gravity on the evolution of
$\delta_c$. In order to study the BAOs, we will assume that the
last term of the equation for the baryonic matter density
contrast, dominates the previous one so that the baryon density
contrast does not grow. This way, it undergoes slowly damped
standing wave oscillations with frequency $c_sk/a$. This is
important because before decoupling, the tightly coupled
baryon-photon fluid oscillates in a similar manner, which refers
to the BAOs. We are going to move to the general relativistic
interpretation of the perturbations, in order to examine the BAOs.
Now we will consider four separate components of the cosmic fluid
so that the total perturbations will be:
\begin{equation}
    \delta\rho = \sum_i \delta\rho_i\, ,
\end{equation}
\begin{equation}
    (\rho + P)V = \sum_i (\rho_i + P_i)V_i\, ,
\end{equation}
\begin{equation}
    \delta P = \sum_i \delta P_i\, ,
\end{equation}
\begin{equation}
    P\Pi = \sum_i P_I\Pi_i\, .
\end{equation}
Firstly Thomson scattering of photons off electrons does not
affect the total energy of the photons, and since we are
considering the tightly coupling approximation for the baryons, it
is taken to prevent photon diffusion, so photons satisfy the
continuity equation:
\begin{equation}
    \dot\delta_\gamma = -\frac{4}{3}kV_\gamma + 4\dot\Phi\, .
\end{equation}
Also baryons satisfy the continuity equation:
\begin{equation}
    \dot\delta_B = -kV_B + 3\dot\Phi\, .
\end{equation}
As far as the number density of protons is everywhere equal to the
number density of electrons, we obtain:
\begin{equation}
    V_B = V_\gamma\, ,
\end{equation}
\begin{equation}
    \frac{\delta(n_B/n_\gamma)}{n_B/n_\gamma} = \delta_B - \frac{3}{4}\delta_\gamma =
    const\, .
\end{equation}
In this approximation, the Euler equation for the baryon photon
fluid will be:
\begin{equation}
    \dot V_\gamma = -a H(1-3\tilde w)V_\gamma - \frac{\dot{\tilde w}}{1+ \tilde w} V_\gamma + \frac{k}{3} \frac{\rho_\gamma}{\rho_B + \frac{4}{3}\rho_\gamma}\delta_\gamma +
    k\Psi\, ,
\end{equation}
where to a good approximation $3\tilde w = \rho_\gamma /
(\rho_\gamma + \rho_B)$ and we also considered $\delta P = \delta
P_\gamma$ which is correct for scales above the Jeans scale.
Lastly, CDM does not interact and has negligible pressure so it
satisfies the continuity and Euler equations respectively:
\begin{equation}
    \dot\delta_c = -kV_c + 3\dot\Phi\, ,
\end{equation}
\begin{equation}
    \dot V_c = -\alpha HV_c + k\Psi\, .
\end{equation}
In this case, we ignore the neutrino perturbation and we get $\Phi
= \Psi$. Now we have to set our initial conditions in order to
work with the equations above. We adopt the adiabatic initial
condition for the density perturbation and by evaluating the 00
component of the Einstein equation with $k=0$ we obtain:
\begin{equation}
    -\zeta = \Phi + \frac{2}{3}\frac{\Psi + (a H)^{-1} \dot
    \Phi}{1+w}\, .
\end{equation}
From this equation with the adiabatic initial conditions, meaning
that $\zeta$ is a constant, and during an era where $w$ is
constant too, we get:
\begin{equation}
    \Phi = \Psi = -\frac{3+3w}{5+3w}\zeta\, ,
\end{equation}
and we note that we have dropped the decaying solution. More
specifically, we set that during radiation domination and before
horizon entry we have:
\begin{equation}
    \Phi = \Psi = -\frac{2}{3}\zeta\, ,
\end{equation}
and
\begin{equation}
    \delta = -2\Phi = \frac{4}{3}\zeta\, .
\end{equation}
Next by combining the continuity and Euler equations we obtain the
equation which describes the acoustic oscillation of the baryon
photon fluid:
\begin{equation}
    \frac{1}{4}\ddot\delta_{\gamma k} + \frac{1}{4}\frac{\dot R}{1+R}\dot\delta_{\gamma k} + \frac{1}{4}k^2c_s^2\delta_{\gamma k} =
    F_k(\eta)\, ,
\end{equation}
where:
\begin{equation}
     F_k(\eta) \equiv -\frac{k^2}{3}\Psi_k(\eta) + \frac{\dot R(\eta)}{1+R(\eta) \dot \Phi_k (\eta)} + \ddot\Phi_k (\eta)\, ,
\end{equation}
\begin{equation}
    c_s^2(\eta)\equiv \frac{\dot P}{\dot\rho}\simeq\frac{\dot P_\gamma}{\dot\rho_\gamma +\dot\rho_B}=\frac{1}{3(1+R(\eta))}\, ,
\end{equation}
\begin{equation}
    R(\eta)\equiv\frac{3}{4}\frac{\rho_B}{\rho_\gamma}\, .
\end{equation}
Thus, we come to the conclusion that the acoustic oscillator is a
forced oscillator, with $F$ being the driving term and also $c_s$
is the speed of sound. The way these equations are structured
reveal that the CDM component evolves without support by any
pressure ($\ddot\delta_c + 2H\dot\delta_c = (3/2)H^2\delta$),
while baryons, at the same time, experience pressure forces
through the term $(k/a)^2c_s^2\delta_B$. This essentially means
that in a MOND Universe, the baryon-photon fluid would oscillate,
almost undamped, until the moment of recombination. Then after the
decoupling, the same oscillations would freeze into the matter
distributions, and the power spectrum would thus be modulated
almost$\sim 100\%$. The observations from BOSS, DESI and SDSS
\cite{BOSS:2016wmc}, indicate that BAOs wiggle $\sim 5-10\%$
hence, this indicates that CDM does not participate in the
oscillations but generates the potential wells
\cite{Dodelson:2011qv}. From the field equations we obtain:
\begin{equation}
    \delta+3\frac{a H}{k}(1+w)V=-\frac{2}{3}\left(\frac{k}{a
    H}\right)^2\Phi\, ,
\end{equation}
\begin{equation}
    \Pi=\left(\frac{k}{a H}\right)^2(\Psi - \Phi)\, .
\end{equation}
These are the constraint equations, and they determine the $F$ in
terms of the perturbations of the neutrinos, the CDM and the
baryon-photon fluid. Due to the fact that neutrinos and CDM vary
slowly, the angular frequency of the driving term is $kc_s$, which
determines the frequency of the oscillation:
\begin{equation}
    \frac{1}{4}\delta_{\gamma k}(\eta)=A_k(\eta) +
    B_k(\eta)\cos{(kr_s(\eta))}+C_k(\eta)\sin{(kr_s(\eta))}\, ,
\end{equation}
where $A_k$, $B_k$ and $C_k$ are slowly varying coefficients.
After matter dominates the evolution, we obtain:
\begin{equation}
    A_k(\eta)=-[1+R(\eta)]\Phi_k\, ,
\end{equation}
where $r_s$ is the distance that sound has had time to travel
since $\eta =0$, and it is called the sound horizon:
\begin{equation}
    r_s(\eta)=\int_0^\eta c_s(\eta)d\eta\, .
\end{equation}
At early times, we consider $R\simeq0$, $c_s=1/\sqrt{3}$ and
$r_s=c_s\eta$. The evolution of $R$ with time will be:
\begin{equation}
    R(\eta)=\frac{3}{4}f_B(1-R_\nu)^{-1}\frac{z_{eq}}{z(\eta)}\, ,
\end{equation}
where $f_B\equiv\rho_B/\rho_m$ and
$R_\nu\equiv\rho_nu/\rho_r=0.40$. Finally for the sound horizon we
obtain:
\begin{equation}
    r_s(\eta)=\frac{2}{3k_{eq}}\sqrt{\frac{6}{R_{eq}}}\ln{\left(\frac{\sqrt{1+R(\eta)}+\sqrt{R(\eta)+R_{eq}}}{1+\sqrt{R_{eq}}}\right)}\,
    .
\end{equation}
At decoupling we find from data that:
\begin{equation}
   R\simeq0.65\, ,
\end{equation}
\begin{equation}
    r_s=1.5\times10^2 Mpc\, .
\end{equation}
In order to get the desired ``snapshot'' of the acoustic
oscillation exactly at the decoupling, we need to define the Silk
damping, which we did not consider above. The Silk damping comes
from the diffusion of the photons, which drag the baryons with
them via the Thomson scattering. In order to calculate the Silk
scale, we first need to examine the Thomson scattering of the
photons off of the electrons. More specifically, Thomson
scattering is essentially Compton scattering, but in this case we
are interested in the regime $T\lesssim10^{-1}MeV$, where the
photonic energy is much less than the electron's mass. Electrons
are considered non-relativistic and thus this process is
classical. Firstly, the electron is at rest and an incoming
photonic plane wave with specific frequency, causes it to
oscillate with that given frequency, and eventually it emits
dipole radiation. In this frame, the Thomson scattering does not
change the photon energy, and the scattered photon has the same
probability of going backwards and forwards, hence its average
momentum is zero. It is important to note that, if the electron
had speed $\upsilon \ll 1$, then the photon energy would also
remain unchanged. Thomson scattering provides us with the opacity:
\begin{equation}
    \tau =n_e\sigma_T a\, ,
\end{equation}
which in turn determines the photon's diffusion length $l_d \sim
\sqrt{N}\lambda_{mfp}$, with $N$ being the number of scatterings.
So the damping scale will be:
\begin{equation}
k_D^{-2} = \int_0^{\eta_{ls}} \frac{1}{6n_e\sigma_T
a}\frac{\dot\tau^2}{\tau^2} d\eta\, .
\end{equation}
Hence, this indicates that the damping continues until the
decoupling. In order to estimate the Silk scale, we assume that
the photons propagates along a random walk in the local baryon
rest frame. The average time between the collisions is $t_c\sim
(n_e\sigma_T)^{-1}$, and the average number of steps during the
time $t$ is $N=t/t_c$, in which a photon diffuses at a distance
$d\sim\sqrt{N}t_c\sim(tt_c)^{1/2}$. So the Silk scale is,
\cite{Planck:2019nip}:
\begin{equation}
    \alpha
    k_D^{-1}\simeq\left(\frac{1}{n_e\sigma_T}\right)^{1/2}\, .
\end{equation}
Before the decoupling and during the matter domination we have
$k_D^{-1}\sim a^{5/4}$, while during the radiation domination
$k_D^{-1}\sim a^{3/2}$. Also every scale $k^{-1}$, is larger than
the Silk scale upon horizon entry, and also the photon diffusion
starts only when $k_S(\eta)=k$. At the decoupling, the Silk scale
is not a well defined concept, due to the fact that at that epoch
$n_e$ falls sharply and $k_D^{-1}\simeq8Mpc$. As a scale, this is
similar to the scale of the photon's mean free path, so $t_c\sim
H^{-1}$ and the kinetic theory is thus no longer valid. Combining
the previous reasoning, we end up with the ``snapshot'' of the
acoustic oscillation at decoupling,
\begin{equation}
    \frac{1}{4}\delta_{\gamma k}=A_k
    +e^{-k^2/k_D^2}[B_k\cos{(kr_s)}+C_k\sin{(kr_s)}]\, ,
\end{equation}
\begin{equation}
    \frac{1}{4}\dot\delta_{\gamma k}=-\frac{k}{3}V_{\gamma
    k}=kc_se^{-k^2/k_D^2}[-B_k\sin{(kr_s)}+C_k\cos{(kr_s)}]\, .
\end{equation}
So at the decoupling $R\simeq0.70$, $r_s\simeq150Mpc$ and
$k_D^{-1}\simeq8Mpc$ and in order to calculate the coefficients,
we assume adiabatic initial conditions and thus we obtain:
\begin{equation}
    \frac{1}{4}\delta_{\gamma
    k}\simeq-(1+R)\Phi_k+\frac{1}{3}e^{-k^2/k_D^2}\cos{(kr_s)}\zeta_k\,
    ,
\end{equation}
\begin{equation}
    \frac{1}{4}\dot\delta_{\gamma k}=-\frac{k}{3}V_{\gamma k}\simeq
    -\frac{1}{3}kc_se^{-k^2/k_D^2}\sin{(kr_s)}\zeta_k\, ,
\end{equation}
where $\Phi_k=-(3/5)T(k)\zeta_k$. The damping term
$e^{-k^2/k_D^2}$, obviously depends on the photon diffusion
length, which depends on the electron number density $n_e$ and the
total expansion history. This is also predicted accurately by the
damping tail of the CMB power spectrum in the $\Lambda$CDM model.


Let us try here to further support the need for DM using the BAOs
and their physics, and thus further clarify more transparently
what we have presented in this section. As we discussed in this
section, baryons and photons were tightly coupled until the
recombination era around $z_{rec}\sim 1100$. Therefore, the small
scale oscillations in the baryonic fluid could not grow for
redshifts $z>z_{rec}$. At large scales, gravity overpowers the
pressure of the baryon-photon fluid and allows perturbations to
grow. The perturbation evolution of the baryonic fluid is governed
by the following evolution equation,
\begin{equation}\label{evolutioneqnbaryon}
\ddot{\delta}_k^{B}+2H\dot{\delta}_k^B+\frac{k^2c_s^2}{a^2}\delta_k^B=4\pi
G\rho_B\delta_k^B+4\pi G \rho_{DM}\delta_k^{DM}\, ,
\end{equation}
where $c_s^2$ is the sound speed of the baryon-photon fluid,
$\delta_k^B=\frac{\delta \rho_B}{\rho_B}$ the perturbation of the
baryonic fluid energy density, $\delta_k^{DM}=\frac{\delta
\rho_{DM}}{\rho_{DM}}$ the perturbation of the DM fluid energy
density, $H$ is the Hubble rate and $a$ is the scale factor of a
flat FRW Universe. Now usually the term $4\pi G\rho_B\delta_k^B$
is ignored, thus the baryonic perturbations are controlled by DM.
In fact, for $a>a_{rec}$, the baryonic perturbations are entirely
controlled by the DM term $4\pi G \rho_{DM}\delta_k^{DM}$. As we
also explained earlier, photons diffuse from high density to low
density regions in the plasma, and they drag the baryons with
them. This process wipes out small scale fluctuations of the
baryon fluid, the process known as Silk damping.

Let us discuss the validity of the linear perturbation theory of
the baryon fluid. When the baryon perturbations are outside the
horizon, the perturbations grow as $\sim a^2$, but when the baryon
perturbations enter the horizon, the DM and baryon perturbations
share the same amplitude. The perturbations oscillate as an
acoustic wave with a decaying amplitude between the horizon
crossing scale factor $a_e$ and the matter-radiation equality
scale factor $a_{eq}$, that is, for $a_e<a<a_{eq}$. For
$a>a_{rec}$, the baryon perturbations grow due to the dominance of
the DM perturbations. So the physical picture is quite simple,
baryons and photons are tightly coupled due to Thomson scattering
in the primordial plasma, and this creates acoustic oscillations
in the baryon-photon plasma. DM drives these oscillations before
recombination and since DM does not interact with radiation, it
collapses, virializes and forms deep gravitational potential wells
before decoupling. After the decoupling, at recombination era, the
baryons fall in the existing DM gravitational potentials and thus
seed the growth of galaxies and galactic clusters. This process is
expected to be non-linear. So in the absence of DM, this
non-linear baryon collapse would require high temperature baryon
clouds to shrink below \cite{padmanabhan2002},
$$R=\frac{t_{cool}}{t_{dyn}}\sim 100\,Kpc\, ,$$
before the baryons are able to cool efficiently to form galaxies.
Apparently this is contrary to logic and observations, thus the
necessity of DM is apparent. Apart from this, and coming back to
the baryon perturbations, the DM fluid is essential for enhancing
the baryonic density contrasts in the Universe and for initiating
the structure formation process.

In the absence of DM, the term $4\pi G \rho_{DM}\delta_k^{DM}$
would be absent in the evolution equation
(\ref{evolutioneqnbaryon}), therefore the baryons would start to
generate perturbations after the decoupling only, due to the
existence of the photon pressure in the baryon-photon fluid before
decoupling. And in the absence of DM, these perturbations would
grow slowly and with very small amplitudes. This would
significantly delay galaxy formation and galactic cluster
formation, which does not track with observations.

Regarding the BAOs, these arise  from the pressure waves in the
baryon-photon fluid before the decoupling, and these waves freeze
at decoupling, forming in parallel very characteristic scales. If
the baryons do not fall in the DM gravitational wells after the
decoupling, the amplitude of the BAOs peaks in the matter
distributions would be very small. These peaks are strongly
correlated with the CMB, and also the galaxy correlation function
and the matter power spectrum induced from the CMB, would look
quite different. So in the absence of DM, the correlations between
large scale structure and the CMB would break and would be
significantly altered. And also overall the BAOs would look quite
different in the absence of DM. These effects cannot be mimicked
by any MOND theory, which shows the importance of the DM fluid.


\section{CMB Anisotropies, Peaks and the Need for DM}

The greatest and incomparable success of the $\Lambda$CDM is the
explanation of the CMB, and no MOND theory can mimic that. In this
section we shall present the challenges that MOND theories face
when it comes explaining the CMB anisotropies and power spectrum.
The CMB anisotropies are essentially the anisotropies of the
photons distribution function in the Universe, with the photons
being the ones which where released by the coupled baryon-photon
fluid at decoupling. Before the decoupling, the photons were in
thermal equilibrium with the baryons via Thomson scattering, and
therefore, they had a black body distribution of momenta
\cite{lyth2009}. Close to the epoch of the decoupling, the
distribution slowly fell off of equilibrium, which lead to an
anisotropy, different for the two polarization states of the
photon. After decoupling, at $z\sim1000$, the inhomogeneous
gravitational field caused the photons to redshift, and that
redshifting generated more anisotropy, leaving though the
polarization unaffected. The anisotropy is examined through a
perturbation in the intensity, which, in turn, corresponds to a
perturbation in the temperature of the black body distribution,
and by two polarization parameters. The polarization of the CMB
provides a firmer test of the standard cosmological model. It
consists of two geometric types: curl-free E-modes, generated by
the scalar density perturbations from the acoustic oscillations,
and divergence-free B-modes, which at linear order are generated
only by tensor perturbations, the so-called primordial
gravitational waves. Such gravitational waves are expected to be
generated during the inflationary regime and the detection of the
B-modes will be the smoking gun signature of the inflationary era.
The E-mode pattern is sourced by the velocity of the oscillating
fluid at the last scattering surface, rather than its density, so
that the pattern is predicted to be out of phase with the
temperature anisotropies. The precise phase relationship has been
confirmed by the CMB experiments \cite{Planck:2019nip}, providing
a very strong verification of the baryon-photon fluid dynamics.
Furthermore, the amplitude of the large-angle E-mode signal from
the reionization, measures the optical depth $\tau$, which comes
consistent with the $\Lambda$CDM timeline of the structure
formation. Any successful theory, including a possible viable
relativistic extension of MOND, must reproduce this specific phase
correlation and of course the full E-mode power spectrum, which
itself points to the existence of CDM. The lensing of these
E-modes into B-modes by large-scale structure, further analyzes
the growth of gravitational potentials over cosmic time, thus
adding another difficulty to any alternative theory. These angular
anisotropies arise for a plethora of reasons, for example, photons
that reach us are redshifted from different directions, and by
different amounts, but also due to a similar phenomenon to this,
where the matter that scattered the photon radiation has a
peculiar velocity. Additionally, photons experience different
redshifting from different gravitational potential wells. If now,
the energy density has an intrinsic inhomogeneity of
$\delta\rho/\rho$, then the temperature will have a fluctuation
$\delta T/T$. Furthermore, processes happening in the path of the
photon might also contribute to temperature fluctuations, as well
as the processes that wipe out the $\delta T/T$, such as the
thermal Sunyaev-Zel'dovich effect. This effect occurs when a
galaxy cluster is in the line of sight of the photons and arises
from the photons that are scattered off the hot cluster gas
throughout the cluster, thus gaining energy. The scale of the
effect is one arc-minute, while distant clusters contribute to the
CMB spectrum on smaller scales, where the primary anisotropies are
damped. We define the photon distribution function in the locally
orthonormal frame and for the multipoles of the brightness
function we obtain,
\begin{equation}
    \Theta(\eta, x, n)=\sum_{lm}(-1)^l\Theta_{lm}(\eta, x)Y_{lm}(n)=\sum_{lm}\Theta_{lm}(\eta,
    x)Y_{lm}(e)\, .
\end{equation}
For $l\ge 2$ the observed multipoles represent the intrinsic CMB
anisotropy and are denoted by $a_{lm}$, which are:
\begin{equation}
    a_{lm}\equiv \Theta_{lm}(\eta_0, x_0)\, .
\end{equation}
In order to examine the spectrum of the CMB anisotropies we have
to see through the stochastic properties of the CMB multipoles.
Specifically it is demanded that:
\begin{equation}
    \langle a_{lm} \rangle=0\, ,
\end{equation}
and
\begin{equation}
    \langle a_{lm},a^*_{l'm'}
    \rangle=\delta_{ll'}\delta_{mm'}C_l\, ,
\end{equation}
where $C_l=\langle |a_{lm}|^2 \rangle=0$ is the spectrum of the
CMB anisotropy and it is defined by:
\begin{equation}
    C(\theta)\equiv \langle \Theta(e_1)\Theta
    (e_2)\rangle=\sum_l\frac{2l+1}{4\pi}C_lP_l(\cos\theta)\, .
\end{equation}
The cosmic variance of $C_l$ will then be:
\begin{equation}
    (\Delta C_l)^2\equiv \langle (|a_{lm}|^2-C_l)^2\rangle= \langle |a_{lm}|^4\rangle-C_l^2=\frac{2}{(2l+1)\Delta
    l}C_l^2\, .
\end{equation}
If we consider only  small angular scales (or large multipoles),
we can work with the flat-sky approximation, so that the
brightness function is defined as:
\begin{equation}
    \Theta(\vec\theta)=\frac{1}{2\pi}\int{ a(\vec l)e^{i\vec\theta \cdot \vec l} d^2\vec l
    }\, .
\end{equation}
Taking into account only the scalar mode of the brightness
function, we then obtain:
\begin{equation}
    \Theta (\eta, x, n)=\sum_l(-i)^l\sqrt{4\pi(2l+1)}Y_{l0}(n)\Theta_l(\eta,
    k)\, ,
\end{equation}
where, in the adiabatic mode, the transfer function
$\Theta_l(k)=T_l(k)\zeta_k$  for the spectrum of the CMB
anisotropy is:
\begin{equation}
    C_l=4\pi\int_0^\infty T_l^2(k)\mathcal
    P_\zeta(k)\frac{dk}{k}\, ,
\end{equation}
where $T_l(k)$ is a transfer function and it embodies all the
complex physics from the early Universe to present day. We have
already discussed that before the recombination era, the
baryon-photon fluid oscillates in the gravitational potential
wells created by CDM, and actually the baryon oscillations prior
to recombination are due to DM itself. The compression phases, and
essentially the odd peaks, are being compressed relative to the
rarefaction phase, which are the even peaks. Mathematically, this
is caused by the following oscillation term $A_k(\eta) =
-[1+R(\eta)]\Phi_k$ we derived above. The ratio of odd-to-even
peak amplitudes is a function of the baryon-to-photon ratio $R$
and the depth of the potential wells $\Phi_k$. More specifically,
the third peak is the most interesting one, because it's height is
suppressed relatively to the second peak, due to the fact that CDM
maintains the gravitational potential during the change from
radiation domination to the matter domination era. In the absence
of CDM, the potentials decay way faster, if they exist at all, and
on the contrary, allow the third peak to rise relatively to the
second. The observational data show precisely that the third peak
is suppressed comparably to the second peak
\cite{Planck:2019nip,Planck:2018vyg}. The CMB anisotropy spectrum
is divided into three parts:
\begin{itemize}
   \item The Sachs-Wolfe Plateau ($l \lesssim 30$): This is the signature of the primordial perturbations on super-horizon
scales at the last scattering.
   \item The Acoustic Peaks ($l \sim 200-1000$): These are a series of
peaks and troughs. The first peak is characteristic of the total
energy density of the Universe. The relative heights of the odd
and even peaks tell us about the baryon density.
   \item The Damping Tail ($l \gtrsim 1000$): As we have already discussed
photons diffuse out of overdensities during the recombination,
exponentially damping fluctuations on small scales. This is the
$e^{-k^2/k_D^2}$ term.
\end{itemize}
In the $\Lambda$CDM model this tail matches the Planck data almost
precisely. This is a great success of the $\Lambda$CDM and it is
incomparable, related to the antagonist theories. MOND theories
cannot easily reproduce that pattern. For MOND theories to
reproduce that, they would need to match the physics of the
recombination and ensure that the growth of structure after the
recombination does not erase the damping. If now we consider the
sudden decoupling approximation, where the baryon-photon fluid
decouples suddenly at a conformal time $\eta_{ls}$, at the last
scattering, an observer right after the last scattering sees no
anisotropy in the CMB. This model assumes recombination happens
instantaneously, which is not true. In reality, the finite
thickness of the last scattering surface ($\Delta z=80$)
introduces a damping of anisotropies on small scales, an effect we
capture with the damping tail. The total CMB anisotropy in this
approximation will be:
\begin{equation}
    \Theta(e)=\left(\frac{1}{4}\delta_\gamma + e\cdot \nu_\gamma
    \right)_{ls}+\Theta_{SW}(e)\, ,
\end{equation}
where $\Theta_{SW}(e)$ is the Sachs-Wolfe contribution:
\begin{equation}
    \Theta
    _{SW}(e)=\Psi_{ls}+\int_{\eta_{ls}}^{\eta_0}\frac{\partial}{\partial\eta}(\Phi+\Psi)d\eta\,
    .
\end{equation}
With that the sudden decoupling approximation becomes:
\begin{equation}
    \Theta(e)=\left[\left(\frac{1}{4}\delta_\gamma+\Psi\right)+e \cdot
    \nu_\gamma\right]_{ls}+\int_{\eta_{ls}}^{\eta_0}\frac{\partial}{\partial\eta}(\Phi+\Psi)\,
    .
\end{equation}
We assume a complete matter domination after decoupling, and
therefore the CMB anisotropy is almost exclusively determined by
the perturbations on the last scattering surface
$\Theta_{SW}(e)=\Phi_{ls}$. Thus we get:
\begin{equation}
    \Theta_l(k)=\left[\frac{1}{4}\delta_\gamma(\eta_{ls},k)+\Phi(\eta_{ls},k)\right]j_l(k\eta_0)+V_\gamma(\eta_{ls},k)j'_l(k\eta_0)\, ,
\end{equation}
where $j_l(k\eta_0)$, is the spherical Bessel function. We can
observe the Sachs-Wolfe plateau here in the regime $l\lesssim30$.
Here the sudden decoupling approximation is almost perfect, due to
the fact that the corresponding scales are well outside the
horizon at the last scattering. Using that $\delta_m=-2\Phi$ and
the adiabatic condition $\delta_\gamma=-(8/3)\Phi$ we arrive to
the Sachs-Wolfe effect:
\begin{equation}
    \Theta(e)=\left(\frac{1}{4}\delta_\gamma+\Phi\right)=\frac{1}{3}\Phi_{ls}=\frac{1}{5}\zeta_{ls}\,
    ,
\end{equation}
for which the spectrum is:
\begin{equation}
    C_l=\frac{4\pi}{25}\int_0^\infty\frac{dk}{k}j_l^2(k\eta_0)\mathcal{P}_\zeta(k) \xrightarrow{}
    l(l+1)C_l=\frac{2\pi}{25}\mathcal{P}_\zeta(l/\eta_0)\, .
\end{equation}
Moving now to the correlation of the acoustic peaks and the CMB
anisotropies, as we have already discussed, the acoustic peaks
represent a snapshot of the acoustic oscillation at the
decoupling. Thus we obtain:
\begin{equation}
    C_l=4\pi\mathcal{P}_\zeta \int_0^\infty
    \left[A_1(k)+A_2\cos{(kr_s)}e^{-k^2/k_D^2}\right]^2j_l^2(k\eta_0)\frac{dk}{k}\,
    .
\end{equation}
As we can see, the entire shape of the $C_l$ curve is a function
of the cosmological parameters ($\Omega_b$, $\Omega_c$,
$\Omega_\Lambda$, $H_0$, $n_s$, $A_s$), this is a novel
characteristic of the $\Lambda$CDM model. And actually, the very
own fact that we can fit this complex curve with a great precision
is an incredible success of the $\Lambda$CDM model
\cite{Planck:2019nip,Planck:2018vyg}.


\section{CDM and Structure Formation}

The pattern of the baryonic acoustic oscillations which is frozen
into the CMB at recombination, forms the initial conditions for
all the subsequent structure formation. The temperature
anisotropies we have discussed earlier, of the order $\delta T/T
\sim 10^{-5}$, are essentially the primordial seeds from which the
galaxies, the clusters, and even the cosmic web eventually are
generated. However, as we already discussed in the previous
section, the baryon-photon fluid could not have evolved into the
structures we observe today, by itself and without DM. The
perturbations of the BAOs could never be strong enough prior to
recombination without DM. More specifically, the cosmological
parameters we discussed earlier, the matter density $\Omega_m$,
the baryon density $\Omega_b$, the Hubble constant $H_0$, and the
amplitude and tilt of the primordial power spectrum $A_s$, $n_s$
respectively, constrained to within a few percent by experiments
like Planck \cite{Planck:2018vyg}, are the initial conditions for
the growth of all the cosmic structure in the Universe
\cite{lyth2009,padmanabhan2002}. We will basically analyze the
formation of the gravitationally bound objects from the primordial
density perturbation, starting with smaller structures and ending
with the larger galaxy clusters. It is important to note at this
point that we have smoothed the above mentioned quantities on a
scale $R$, which removes structure on scales $\lesssim R$.
Smoothing is important in CMB-related physics. The density
contrast smoothed on scale $R$ is:
\begin{equation}
    \delta_k(z)=\frac{2}{5} \widetilde W(kR)\left(\frac{k}{H_0}\right)^2\frac{T(k)}{\Omega_m}\frac{g(z)}{1+z}
    \zeta_k\, .
\end{equation}
A quite important outcome of structure formation in the
$\Lambda$CDM is that it reproduces the successes of MOND on
galactic scales. Cosmological simulations indicate the deep
correlation between the observed baryonic mass to the centripetal
acceleration, which MOND theories are engineered to explain. Also
$\Lambda$CDM simulations show that the Milgrom's constant
$\alpha_0$ is not universal, something that contradicts the MOND's
assumption that $\alpha_0$ is an epoch independent constant
\cite{Mayer:2022qhk}. Particularly, these simulations show that
$\alpha_0$ should be increasing with redshift, due to the coupling
between baryons and DM that evolves during structure formation. We
find that the linear regime ends at the epoch $z_{nl}$, which we
obtain from $\sigma(R, z_{nl})=1$, where $\sigma(R, z)$ is the rms
density contrast:
\begin{equation}
    1+z_{nl}=\sigma_0(R)\, ,
\end{equation}
and $\sigma_0$ is calculated at the present epoch:
\begin{equation}
    \sigma_0^2(R)=\int\widetilde W^2(kR)\mathcal P_0(k)\frac{dk}{k} \sim
    \mathcal{P}_0(R^{-1})\, .
\end{equation}
We can see that the spectrum $\widetilde W^2(kR)\mathcal{P}_0(k)$
of the smoothed density contrast, has a peak at $k\sim1/R$. This
position of the peak indicate that an overdense region has size
$R$. So when the linear regime comes to an end, such regions
typically collapse, due to the fact that the inward peculiar
velocity at its edge, is at same order as the Hubble velocity.
This means also that, due to the fact that the overdense regions
account for half of the volume, a significant amount of the matter
of the Universe collapses to structures. The density contrast is
still evolving linearly, which essentially means that while some
regions collapse, bigger ones expand with the Universe. At
$z\sim10$ to $30$, the Jeans mass is of order $10^4M_\odot$, which
means that the lighter objects will form earlier and they will be
entirely of DM, because they are below the baryon Jeans mass. From
the virial theorem we know that when an overdense region
collapses, it reaches the virial equilibrium at a radius almost
half its maximum expansion radius. For the spherical top-hat
perturbation, the final overdensity relative to the background at
virialization would be of the order $\Delta_{vir} \approx 178$ in
an Einstein-de Sitter Universe, which defines the boundary for DM
halos, and therefore determines the structure formation. Among
objects of the same mass, the ones that form earlier have a higher
virial velocity because they are more compact, due to the fact
that the Universe, upon their structure, was smaller and denser.
Galaxies have virial velocities of the order $100$km$s^{-1}$,
while clusters, $1000$km$s^{-1}$, an order higher. Here we assume
that the collapsing object loses no energy, so it applies to the
CDM. The angular momentum of the CDM and baryons is conserved
respectively and individually. A galaxy will be a spherical and
static distribution of DM (dark halo), within which there is a
more tightly bound distribution of baryons, due to the fact that
baryons can radiate energy in order to become more tightly bound,
mostly in stars.

The non-linear evolution of structure is studied thoroughly with
N-body simulations and indicate that the DM halo mass function is
described by the press-Schechter formalism. The study of this
evolution sets the basis for the understanding of galaxy
clustering, as well as the bias factor $b$ \cite{lyth2009}. The
total mass density perturbation can be mapped through weak
gravitational lensing from the distant galaxies. As we have
already pointed out, on large scales, the DM density perturbation
evolves linearly, while on the small scales, the evolution is
determined through gravitational N-body simulations. The evolution
of the baryonic structures, on the other hand, is followed through
hydrodynamical N-body simulations. Ignoring the properties of each
galaxy, it is possible to map the large distribution of DM through
the galaxies' clustering, directly from the dark halos. If we
consider a Gaussian primordial density perturbation we obtain:
\begin{equation}
\mathcal{P}_g(k,z)=b^2\mathcal{P}_\delta(k,z)\, ,
\end{equation}
where $\mathcal{P}_g$ is the power spectrum of galaxies,
$\mathcal{P}_\delta$ is the power spectrum of matter and $b$ is a
bias factor, which is expected to be constant for a specific epoch
and a well defined class of galaxies. Even if the primordial
density perturbation is considered being a Gaussian, we must also
consider the effect of primordial non-Gaussianity. The primordial
bispectrum for local non-Gaussianity is specified by the parameter
$f_{NL}$. So the power spectrum relations becomes:
\begin{equation}
    \mathcal{P}_g(k)=(b^2+\Delta
    b(k)f_{NL})\mathcal{P}_\delta(k)\, ,
\end{equation}
where
\begin{equation}
    \Delta
    b(k)=2(b-p)\delta_c\frac{\Phi_k}{\delta_k}=\frac{3(b-p)\delta_c\Omega_mH_0^2(1+z)}{k^2T(k)g(z)}\,
    .
\end{equation}
This way, the galaxy distribution can give constraints on
primordial non-Gaussianity, that have comparable weighting to the
ones from the CMB anisotropy, with different length scale. The
structure formation discussed in this section cannot easily be
generated by any MOND theory, without adding a missing mass
component, perhaps neutrinos. And still, even additional mass
components are added in MOND theories, the details of structure
formation predicted by the $\Lambda$CDM is incomparable.

\section{Direct Comparison of MOND Theories and the $\Lambda$CDM Model}

Having discussed the great and incomparable successes of the
$\Lambda$CDM, in this section we wrap these up and we additionally
present evidence that contradict the MOND paradigm. But caution is
needed, MOND is mostly successful at galactic scales, and fails at
cluster and cosmological scales, and sometimes even at galactic
scales, when it comes to dark galaxies. So we discuss here in
organized form the failures of MOND theories and the successes of
the $\Lambda$CDM paradigm. The key problems we present here are:
the damping problem of the BAOs, the CMB peak ratio problem, the
timescale problem on structure formation and the problem regarding
the baryon fraction in clusters.

\subsection{BAO Damping Problem}

As we discussed earlier, from the evolution of baryon
oscillations, the baryons include a pressure term in their
evolution perturbation equations, and CDM does not have such a
term. Before the recombination era, the baryonic oscillations
occurred dominantly due to DM, and after the decoupling the
oscillations froze into the matter distributions created by DM
gravitational waves. The observations of BOSS
\cite{BOSS:2016wmc,Dodelson:2011qv} indicate damping, which
requires that $85\%$ of the matter does not participate in the
matter oscillations. Therefore the majority of the matter helps
the oscillations to be generated, but does not actually
participate in them, and this matter is DM. To be more specific,
the BAO feature in the galactic correlation function indicates an
amplitude of $\sim 5\%$ compared to the smooth broadband power,
which corresponds to the density parameter $\Omega_c h^2 \approx
0.12$ from the Planck data. In order for DM to explain this
feature, it should be required that some mechanism damps the
oscillation without the presence of any matter component, or it
should be required that baryons themselves erase their
oscillations. If one considers a purely baryonic Universe, the BAO
wiggles would have an amplitude  $\sim100\%$ and there is no such
MOND theory that can actually preserve the damping tail which is
observed in the CMB \cite{Dodelson:2011qv}.

\subsection{CMB Peak Ratio Problem}

In a previous section we discussed the CMB anisotropies and we
showed that the depth of the gravitational potential wells
$\Phi_k$ formed by DM depends strongly on the presence of DM. It
is DM itself that maintains $\Phi_k$ and that it remains
decay-free during radiation domination, and when the modes
corresponding to the third CMB peak enter the Hubble horizon. From
the CMB experiments \cite{Planck:2019nip,Planck:2018vyg} it is
observed that the third peak of the CMB is suppressed in
comparison to the second peak, a direct confirmation of DM.
Specifically, the exact measurements indicate that the ratio of
the third over the second peak is
\cite{Planck:2019nip,Planck:2018vyg},
\begin{equation}
    \frac{P_3}{P_2}=0.97 \pm 0.0011\, ,
\end{equation}
and in a Universe without DM and with only baryons ones obtains
\cite{McGaugh:1999cd},
\begin{equation}\label{nem}
    \frac{P_3}{P_2}\sim 0.51\, ,
\end{equation}
so the discrepancy is notable. The acoustic-peak ratio (\ref{nem})
represents a baryonic limit of the perturbations we consider in
the present section. The Planck references
\cite{Planck:2019nip,Planck:2018vyg} provide the observational
results to be compared with the theoretical derivation (\ref{nem})
which is a result which obtained by using the physics described in
Ref. \cite{Hu:2001bc}, see \cite{McGaugh:1999cd} for full
details\footnote{Actually Ref. \cite{McGaugh:1999cd} gives the
opposite fraction $\frac{P_2}{P_3}=1.88$}.

Apparently, MOND theories cannot explain these measurements, and
MOND theories rely on using missing mass components, like
neutrinos. Thus, the very own problem that allegedly MOND theories
solve, that is, the lack of any additional mass components save
the baryons, is again reintroduced in another form. It is not DM,
it is missing neutrinos.

\subsection{Structure Formation Problem}

In the $\Lambda$CDM Model, there is strict hierarchical structure
formation, thus the smaller structures collapse earlier, and DM
haloes form first. This leads to a characteristic mass function
and thus a specific epoch of reionization at redshift $z\sim
10-12$ occurs. The MOND description leads however on faster growth
in structure formation and to two predictions
\cite{Mayer:2022qhk}: massive galactic clusters at high redshifts
are overproduced and there is excessive matter power spectrum at
smaller scales, thus leading to the overproduction of dwarf
galaxies.  Thus MOND fails at the galactic clusters scales and the
matter power spectrum \cite{Mayer:2022qhk}.

Also, the baryon fraction in galactic clusters in the presence of
DM is \cite{padmanabhan2002,Planck:2018vyg},
\begin{equation}
    f_b \equiv \frac{\Omega_b}{\Omega_m} \approx 0.16
\end{equation}
and X-ray observations indicate that $f_b \sim 0.15$ in massive
galactic clusters \cite{padmanabhan2002}. Using gravitational
lensing, we do understand why MOND fails in predicting the above
baryon fractions. MOND requires again invisible matter like
sterile neutrinos to explain the above, and thus the same problem
of requiring missing mass in MOND theories arises.

\subsection{Gravitational Lensing from Galaxies}

The gravitational lensing is one of the physical phenomena that
MOND theories cannot describe consistently at all, without adding
missing invisible matter. Gravitational lensing occurs when light
from distant sources is bent due to the gravitational field of a
foreground object, which acts as a lens. When galaxies act as
gravitational lenses it is important to decompose its total mass
to several components, that is the DM halo and the baryonic
matter, which is comprised by gas and stars
\cite{Bartelmann:2010fz}. Hence, in the presence of DM in the
Universe, massive galaxies should be filled with thousands of
low-mass subhalos. There exist various galactic gravitational
lenses, and the important cases are listed below:
\begin{itemize}
    \item Early type galaxies: These galaxies are massive
    elliptical and lenticular galaxies according to the Hubble
    classification and these produce strong lenses.

    \item Late type galaxies: these are spirals and constitute a
    rarer form of gravity lenses, due to the fact that their total
    mass is diffuse and contain dust.

    \item Dwarf galaxies: these are small galaxies, hence their
    name, and are mostly irregular in which the young stars are
    not contained in a spiral.
\end{itemize}
Now the question is how the $\Lambda$CDM is connected to the
gravitational lenses? Gravitational lenses actually prove the CDM
part of the $\Lambda$CDM model. From the observations of
gravitational lenses, it was proven that the mass of the lensing
galaxies is way larger than the visible mass coming from the stars
of the lensing galaxy. This is a direct non-dynamical evidence
that DM forms the gravitational wells where galaxies
hierarchically virialize their baryonic matter. The visible light
traces the baryonic matter, but the gravitational lenses point out
the total mass, which reveals in a non-dynamical way the presence
of the DM halo \cite{Treu:2010uj}. Also the gravity lenses probe
the DM subhalo structure of the lensing galaxies. The $\Lambda$CDM
itself predicts that DM subhalos exist in galaxies, and in strong
lensing galaxies there exist flux ratio anomalies, which are
explained by the gravity perturbations of small DM subhalos
\cite{Mao:1997ek}. These anomalies are direct evidence of an
overall abundance of low mass DM subhalos, since the anomalies
detect failed star formation in these subhalos. This basically
constitutes the missing satellites problem of the $\Lambda$CDM,
which is solved nowadays \cite{Kim:2017iwr}.

Now coming back to the gravitational lenses, they play an
important role on measuring the matter distribution. These
measures yield the amplitude of the matter clustering $\sigma_8$,
the total matter density $\Omega_m$, and the $\Lambda$CDM predicts
$S_8=\sigma_8 \sqrt{\Omega_m /0.3}$, while the latest lensing
surveys indicate a value of $S_8$, which is about $8-10\%$ lower
than the current value obtained from CMB
\cite{DiValentino:2020vvd}.


\subsection{Direct Evidence of DM: Dark Galaxies, and Notable Galactic Clusters}

MOND theories cannot explain at all the existence of dark galaxies
and systems that are ultra diffuse galaxies (UDGs). Dark galaxies
are essentially galaxies which contain very few stars, and thus
cannot be spotted by optical telescopes. The mass of these
galaxies and the hydrodynamical equilibrium is attributed in DM,
and most of these galaxies are traceable due to their clouds of
neutral hydrogen gas, which is spotted using radiotelescopes. Many
studies indicate \cite{Anand:2025czy} that dark galaxies are
basically failed ordinary galaxies, formed in the clumps of DM,
but their gas failed to collapse enough in order to start enough
star formation, or their gas was stripped from them due to some
dynamical reason, such as a neighboring galaxy. There exist
several well known dark galaxies, such as Cloud 9
\cite{Anand:2025czy,Benitez-Llambay:2023kyu}, AGC 114905
\cite{Afruni:2025fmx}, the Dragonfly 44 \cite{vanDokkum:2016uwg},
NGC1052-DF2 \cite{vanDokkum:2018vup,Buzzo:2025zmu}. The dark
galaxies strongly challenge the MOND paradigm, since MOND cannot
explain at all the hydrodynamical equilibrium of starless systems.
Also MOND theories cannot explain at all the formation of DM halos
at subgalactic mass scales.

Now let us discuss two clusters of particular interest related to
the nature of DM itself, the Bullet cluster and the Abell 520
cluster. The Bullet cluster occurred from the collision of two
massive clusters at speeds $\sim 5000$ km/s. This cluster is the
main gatekeeper of DM and in fact of collisionless DM, since the
DM halo passed unaffected by the collision of the original
clusters. Although there exist studies
\cite{Zhao:2007mz,Angus:2007qj}, which indicate that MOND is able
to produce the offset of the gravity lens signal, in the same
study it was discussed that the Bullet cluster demands a
collisionless in nature, non-baryonic component, in order to fully
agree with the data. Thus the MOND perspective fails.

On the other hand, the Abell 520 cluster
\cite{Mahdavi:2007yp,Mahdavi:2012zy,Peel:2017liv} indicates that
DM shows a collisional nature, although this is debatable
\cite{Peel:2017liv}. These two clusters indicate an interesting
feature of DM, the fact that it can behave in some cases as
collisional and in other cases as collisionless. This is quite
interesting regarding the aims of this article which is to show
that DM itself might have a scale dependent EoS, and that the MOND
postulates are not elements of a theory that describes galactic
scales, but manifestations of DM characteristics at galactic
scales. We discuss this perspective in a later section in a more
focused way.

We need to stress that the Bullet Cluster and Abell 520 provide
two interesting examples for DM dynamical behavior, but these two
examples are far from being equivalent, concerning the overall
distribution of DM and of the baryonic matter in merging systems.
In the case of the Bullet cluster, the X-rays emitted by
inter-cluster gas provide a clean cut example of the fact that the
dominant gravitating component does not coincide with the
collisional baryonic gas. On the other hand, Abell 520 involves a
far more complex merger environment, and thus the DM concentration
interpretation is more vague, compared to the well studied Bullet
cluster case. Hence the two systems cannot be regarded as being
equally observationally strong regarding their actual
observational behavior. The Abell 520 example is currently
debated, so it acts as motivation for further analysis of the
nature of DM and how DM behaves in the presence of strong and
unstable dynamical environments and at different scales.

\section{The Spin Problem of Galaxies and DM}

Galaxies are formed within DM halos and at first the gas that will
later form the galaxy's stars, has the same spin as the DM halo.
As the gas cools down, we expect it to fall inwards and form a
rotating disk. However, simulations that were run, showed that the
gas cooled too quickly and lost a large amount of its angular
momentum to the DM halo, through dynamical friction
\cite{1991ApJ...380..320N}. Observations though, do not match
these simulations, since we observe larger disks that rotate
faster than predicted. This discrepancy is known as the spin
problem or the angular momentum catastrophe. In order to solve
this problem in the $\Lambda$CDM framework spin segregation was
developed \cite{Fall:1980sj}. Both baryonic gas matter and DM are
pulled into a gravitational potential well, but because of the
different physical natures of the two, they behave differently. DM
as it falls into the potential, it conserves its individual
angular momentum, if it is collisionless in nature. Baryonic
matter, on the other hand, is collisional, so through collisions,
shocks and radiation, the gas loses energy, falls deeper into the
potential well and transfers its angular momentum. Spin
segregation works as follows: At first, both gas and DM have
relatively high angular momentum. As the gas falls in, it gets
heated through accretion shock, and afterwards cools down through
X-ray emission. By losing energy, it also loses pressure support,
and therefore sinks towards the center of the DM halo. While the
sinking gas is still rotating, it transfers some of its angular
momentum, through dynamical friction, a purely gravitational
process, to the DM. As a result, DM ends up with the original
angular momentum of the falling material, and the gas now, as it
settles into a rotating disk, has a different angular momentum
profile from DM. Within the $\Lambda$CDM paradigm, the spin
problem was largely solved through more complex baryonic physics
implementations. More specifically, spin segregation is triggered
by important mechanisms, depending on the scale of the problem:
supernovae feedback, and active galactic nuclei feedback
\cite{Brook:2012rr,Fabian:2012xr}. These are essentially the tools
that allow baryons to conserve their spin and form the disks we
observe. Regarding supernovae feedback it comes from dying massive
stars with masses $>8M_\odot$, which explode and transfer their
kinetic energy. The spatial scale of supernovae feedback is
sub-kiloparsec and it affects the interstellar medium in the
neighborhood of star forming regions. Its role in the process of
spin segregation is to prevent the gas from becoming too
concentrated at the early stages, so it preserves its angular
momentum for disk formation. Although only $\sim 10\%$ of the
total energy ($\sim 10^{51}$ erg) released by a supernova couples
to the gas, if we have multiple supernovae in a region where there
is star formation, they jointly drive very powerful galactic
winds. The galactic winds physically remove low angular momentum
material that has already moved to the center. Additionally, they
transfer turbulence and thermal energy to the remaining gas, so it
does not cool and collapse into a low spin core. Through this
exact process the galaxy retains its high spin gas. Regarding
active galactic nuclei feedback, it comes from supermassive black
holes with masses $10^6-10^{10} M_\odot$ that accrete matter and
convert the gravitational potential energy of the infalling gas to
radiation. Its spatial scale extends from the galactic nucleus to
beyond the circumgalactic medium. Its primary role in spin
segregation is preventing the cooling of the gas in massive halos.
Active galactic nuclei feedback works two ways. One way is quasar
mode, in which outflows coming from wide angles remove gas from
the entire galaxy, so star formation cannot occur. The other way
is radio mode, where bubbles of hot plasma are created in the
circumgalactic medium from relativistic jets, preventing it from
cooling and therefore also preventing low angular momentum gas
from moving in to the galaxy.

It is important to note that these two mechanisms work
sequentially and complementary. Initially, supernovae feedback
dominates. In this period the gas puffs up so over-cooling is
prevented, while also outflows remove material with low spin,
allowing the gas to preserve its angular momentum. Second, during
the formation of the disk, supernovae feedback still regulates
star formation. During accretion the mass of the galaxy builds up,
until when it reaches a point where supernovae feedback is no
longer effective ($M_{\star} \gtrsim 10^{10} M_\odot$). At this
point active galactic nuclei feedback starts dominating. That
prevents cooling and cuts off the supply of low angular momentum
gas. Cosmological simulations that include these feedback
processes, can now reproduce the observed angular momentum of
galaxies of different masses, resolving, essentially, the spin
problem. The spin problem constitutes a challenge for baryonic
physics within the $\Lambda$CDM regime and not for DM
individually. As we discussed, spin segregation does not demand
any modifications to gravity. MOND theories do not offer any
natural process that explains the angular momentum of galaxies we
observe, as it also removes the DM halo which is crucial for
retaining angular momentum and transferring dynamical friction. So
in MOND theories, galaxies would have even lower angular momentum
than the initial $\Lambda$CDM simulations, and therefore would
worsen the problem rather than provide a solution.

Apart from that, usually the angular momentum of a galaxy is
expressed in terms of the dimensionless parameter,
$$\lambda=\frac{L\,E^{1/2}}{G\,M^{5/2}}\, ,$$
where $L$ is the spin of the galaxy, $E$ its energy and $M$ its
total mass (baryonic and DM). Disk galaxies observations suggest
that $\lambda\sim 0.4-0.5$. In the absence of a DM halo, assuming
that the protogalaxy is an self-gravitating cloud of baryon gas,
the binding energy of the protogalaxy is $E\sim \frac{GM}{R}$ and
since $M$ remains  constant during the collapse $E\sim
\frac{1}{R}$, so the spin parameter is $\lambda\sim R^{-1/2}$. The
gas cloud has to collapse by a factor
$\left(\frac{\lambda_d}{\lambda_i}\right)^2\sim 400$ before it can
spin up enough to form a rotationally supported system, where
$\lambda_i$ is the initial value of the spin parameter $\lambda$
produced by the tidal torques. Hence, in order to form a
rotationally supported disk of mass $M\sim 10^{11}M_{\odot}$ and
radius $10$Kpc, the baryonic matter needs to collapse by an
initial radius of 4Kpc, and this whole process would take a time
amount of the order $t_{coll}=\frac{\pi}{2}\left(\frac{R^3}{2GM}
\right)^{1/2}$ \cite{padmanabhan2002}, which is $t_{coll}\sim
5.3\times 10^{10}$years. This amount of time is larger than the
age of the Universe. This simple example shows that without DM,
very simple problems of galactic dynamics turn out to yield
nonsensical results. This indicates that the MOND paradigm simply
reproduces the dynamics of the rotation curves and the Baryonic
Tully-Fisher relation, but fails at a fundamental level to
reproduce physically consistent result, at galactic, supergalactic
and cosmological scales. Thus we are led to the question, if MOND
is a successful description of galactic level dynamics, why not
using MOND as a phenomenological description of nature at galactic
scales, instead of a theory that explains the galactic scale
dynamics? This is considered in the next section.

\section{The Perspective of Scale-dependent DM, its
Realization via Scale Dependent DM and the MOND Postulates}

MOND theory is based on several postulates that seem to
successfully describe the galactic dynamics in terms of rotation
curves, but MOND fails utterly in describing consistently the
Universe at cluster, supercluster and cosmological scales. Several
recent fails of MOND theories can be found in Refs.
\cite{mondfail1,mondfail2,Russell:2026jqw,Pittordis:2025wxb}, see
also \cite{Desmond:2025pmk} for the successes and fails of MOND.

Taking into account the fact that MOND offers some interesting
perspectives for galactic dynamics and specifically the rotation
curves of some galaxies, we aim here to view MOND not as a theory
which is behind the physics of galaxies, but as a phenomenological
description of DM at galactic scales. So MOND in our perspective,
is not the solution to the problem of missing matter, it is the
description of DM at galactic scales. So the properties of MOND
will be the properties of DM at galactic scales. Since DM behaves
as collisionless at cosmological scales, while at cluster scales
behaves as collisionless and as collisional, we aim here to model
DM as having a scale dependent EoS which can accommodate such
behaviors. Hence, MOND is merely a description of how
scale-dependent EoS DM should behave at galactic scales.

Let us analyze in brief our proposal, MOND is based on altering of
Newton's second law at low accelerations, by introducing the
interpolation function $\mu(x)$ which behaves as,
\begin{equation}
a\,\mu\!\left(\frac{a}{a_0}\right) = \frac{GM_b(r)}{r^2}\, ,
\label{eq:mond}
\end{equation}
with $a_0$ being the characteristic MOND acceleration scale and
also $M_b(r)$ is the enclosed baryonic mass. Taking the deep-MOND
limit ($a\ll a_0$), Eq.~(\ref{eq:mond}) gives,
\begin{equation}
a = \sqrt{a_0\,\frac{GM_b(r)}{r^2}} \;\;\Rightarrow\;\; v_\phi^4 =
G\,M_b\,a_0, \label{eq:btfr_mond}
\end{equation}
and thus the empirical baryonic Tully-Fisher relation is
reproduced,
\begin{equation}
M_b \sim v_\phi^4,
\end{equation}
which naturally explains the flattening of the rotation curves
without invoking any DM.

On the contrary, scale-dependent EoS DM introduced in Refs.
\cite{Oikonomou:2026vkp,Oikonomou:2025bsi}, is based on the two
disciplines, Newton's gravity and DM has a scale-dependent EoS of
the form,
\begin{equation}
P(r) = K(r)\,\rho(r)^{\gamma(r)}, \label{eq:eos_variable}
\end{equation}
with $\gamma(r)$ and $K(r)$ describing the local variations in the
effective compressibility and the entropy of the essentially
collisional DM fluid. This system satisfies of course hydrostatic
equilibrium,
\begin{equation}
\frac{dP}{dr} = -\rho(r)\,\frac{G\,M(r)}{r^2}, \label{eq:hydro_eq}
\end{equation}
with $M(r)$ being the total enclosed mass, mostly comprised by DM.
The circular velocity is defined as follows,
\begin{equation}
v_\phi^2(r) = \frac{G\,M(r)}{r}. \label{eq:v_circ}
\end{equation}
In the specific class of models analyzed in Ref.
\cite{Oikonomou:2025bsi}, both the baryonic and canonical
Tully-Fisher relations were reproduced for galaxies which are DM
dominated or relatively young and medium mass stable disk systems.
The model of Ref. \cite{Oikonomou:2025bsi} relied on a
parametrization of the polytropic index as follows,
\begin{equation}\label{above}
\gamma(r) = \gamma_0 - \delta_\gamma\,\tanh\!\left(\frac{r -
r_\gamma}{2.0\mathrm{Kpc}}\right),
\end{equation}
with the best-fit parameters which provide compatibility with the
SPARC data, for galaxies exhibiting an optimal fit and flat
rotation curves, being,
\begin{equation}\label{choice}
 \gamma_0 = 1.0001, \qquad
\delta_\gamma = 1.2\times10^{-9}\, .
\end{equation}
The parameters $\gamma_0$ and $\delta_\gamma$ which enter the
function $\gamma(r)$ which enters the scale-dependent EoS must be
distinguished as parameters that play a crucial role for
distinguishing the behavior of the 175 SPARC galaxies studied in
detail in Ref. \cite{Oikonomou:2025bsi}. The choice of values made
above is important in order to ensure a MOND-like behavior at
large radii from the cores, but the choice of values made in Eq.
(\ref{choice}) is the same for all the 175 SPARC galaxies. So the
values (\ref{choice}) determine the general functional form of
$\gamma(r)$, but these are not fit parameters. The actual
parameters that are fitted are $K_0$ and the central density of
each galaxy, and these vary from galaxy to galaxy. Hence,
$\gamma_0$ and $\delta_\gamma$ characterize a common to all
galaxies functional form of $\gamma(r)$, and the galaxy specific
parameters are $K_0$ and the central density $\rho_0$. Therefore
we cannot quote observational constraints on the parameters
$\gamma_0$ and $\delta_\gamma$, these are common for every of the
175 SPARC galaxies studied in Ref. \cite{Oikonomou:2025bsi}, and
these values provide a general functional form for $\gamma(r)$
which yields a viable phenomenology for each of the viable
galaxies when $K_0$ and $\rho_0$ are fitted. Hence, $\gamma_0$ and
$\delta_\gamma$ are not newly introduced optimized parameters for
the current analysis. This is an interesting behavior since the
EoS is a nearly isothermal at the core and inner halo regions
since $\gamma \simeq 1$, however the important feature of this
model is the large distance behavior. Let us show this, so by
taking the isothermal limit, Eq.~(\ref{eq:eos_variable}) reduces
to,
\begin{equation}
P = K_0\,\rho,
\end{equation}
where $K_0$ represents the effective entropy constant. As it
proves, the important feature is not the nearly isothermal
behavior, but the fact that at large radii, the entropy is
approximated by the constant $K_0$. This is vital, as we show now.
Using Eqs.~(\ref{eq:hydro_eq}) and (\ref{eq:eos_variable}) in this
limit we obtain,
\begin{equation}
\frac{1}{\rho}\frac{d\rho}{dr} = -\frac{G\,M(r)}{K_0\,r^2}\, .
\end{equation}
For the quasi-isothermal sphere we assumed above, we obtain
approximately $\rho(r)\sim r^{-2}$, hence we obtain,
\begin{equation}
M(r)\sim r, \quad \rightarrow \quad v_\phi^2(r)=\frac{GM(r)}{r} =
\text{constant}.
\end{equation}
Therefore, the flat rotation curves can emerge naturally  in the
framework of scale-dependent self-interacting DM, by using
thermodynamic equilibrium and also that the entropy function
behaves asymptotically as $K(r)\sim K_0$ and in fact that the
entropy function is nearly constant for all radii. The latter
affects the medium and large radii, but not the radii near the
core, where the scale dependent $K(r)$ offers an insightful
solution to the cusp-core problem. Thus in our perspective, the
deep MOND limit of Eq. (\ref{eq:btfr_mond}) should not be viewed
as a prediction, but it should viewed as the desirable behavior of
self-interacting DM at large radii from the core. The MOND the
asymptotic scaling, is perfectly reproduced by scale-dependent
self-interacting DM. We should stress here that the small
numerical value used for the parameter $\delta_\gamma$ must not be
interpreted that it implies a precise observational determination
at the $10^{-9}$ level. It rather characterizes the functional
form of $\gamma(r)$ and it is thus a parameter that yields an
interesting phenomenological behavior for all viable galaxies at
large radii from the core. The central result of this section does
not dependent on whether this parameter is statistically
significant across the 175 SPARC galaxies. Our aim what to show
that the flat rotation curve behavior for galaxies occurs in the
isothermal limit,
\begin{equation}
\gamma(r)\to 1\, ,
\end{equation}
Thus, the hydrostatic equilibrium then yields the asymptotic
density profile,
\begin{equation}
\rho(r)\sim r^{-2}\, ,
\end{equation}
and therefore,
\begin{equation}
4\pi\int_0^r \rho(r')r'^2,dr' \propto r\, ,
\end{equation}
from which it follows directly,
\begin{equation}
\frac{GM(r)}{r} \simeq {\rm constant}\, .
\end{equation}
The behavior above is an essential mathematical behavior which
underlines the MOND behavior at galactic scales.

Before closing this section we need to note that the
scale-dependent behavior of the effective dark sector highlighted
in this section is not meant to hold true when extrapolating to
arbitrarily large scales. This is an interesting perspective to
study separately though because if DM is a sort of mirror DM it
could behave like this. Indeed, if DM contains atoms, elementary
particles, it could be behave differently at galactic scales and
differently at cluster and cosmological scales. Thus the physical
scenario we analyzed in this section requires an effective
collisional and thermodynamic behavior which is responsible for
the galactic scale phenomenology, but this behavior would be
different at cluster and cosmological scales, so that DM behaves
as collisional at larger scales, at least in most of the cases,
since the Abell 520 shows a behavior of collisional DM. Thus the
scale-dependent behavior affects the behavior of DM, depending
also on the composition of the DM and its behavior in different
environments. Thus in this work with our new perspective for
viewing MOND theories, we did not attempt to perform a cluster
scale fit or a transition from galactic to larger scales. This
task is an interesting research line which we aim to address in a
future work, that is, going to study the behavior of scale
dependent EoS at cluster and cosmological scales. Hence this work
was intended to function as a proof-of-concept reinterpretation of
an existing fit, and not as a new quantitative result. Also we
need to note that the construction presented in this section is
related to a much broader class of self-interacting DM, in the
context of which the effective behavior of the DM sector differs
between the galactic and cluster environments. Specifically, this
said behavior that DM exhibits non-negligible self-interactions at
galactic scales and remains essentially collisionless at (most)
cluster scales is also met in velocity-dependent self-interacting
DM models, see Refs.
\cite{Wandelt:2000ad,Spergel:1999mh,Tulin:2017ara} for a fully
detailed analysis of these perspectives. Our approach differs from
the context of the velocity dependent self-interacting DM, in
which the transition is parameterized by velocity dependent
self-interaction cross sections,
\begin{equation}
\sigma=\sigma(v)\, ,
\end{equation}
but in our approach, the effective galactic-scale behavior is
achieved via a scale-dependent EoS,
\begin{equation}
P(r)=K(r)\rho(r)^{\gamma(r)}\, .
\end{equation}
The two phenomenological approach lead to qualitatively similar
phenomenology at galactic scales, but the present scale-dependent
EoS concept is very new and its full analysis is lacking in the
literature, so we hope to address some aspects not covered in the
literature, such as the transition from galactic to larger scales.

\section{Conclusions}

In this article we thoroughly examined the arguments that validate
the $\Lambda$CDM Model and demonstrated its strength and
superiority over the MOND theories. More specifically, we mainly
analyzed the three main pillars of modern cosmology: the baryon
acoustic oscillations, the CMB anisotropies and structure
formation, through the lens of cosmological perturbation theory.
We came to the known conclusion that the provided data from
multiple experiments require cosmologically a collisionless,
invisible and more importantly non-baryonic form of matter.
Analytically we obtained that the damping of BAO wiggles we
observe, demands that about $85\%$ of the total matter does not
take part on the oscillations happening, which is something
naturally explained by CDM \cite{Dodelson:2011qv,BOSS:2016wmc}.
Furthermore, the suppression of the third peak relative to the
second in the CMB anisotropy power spectrum, shows that CDM
potential wells should be maintained during radiation domination
\cite{Planck:2019nip,Planck:2018vyg}. The hierarchical growth of
structure in the $\Lambda$CDM Model matches the observations of
galaxy distribution and also reproduces the successes of MOND on
galactic scales. Apart from these cosmological tests, direct
observations of dark galaxies like Cloud 9, AGC 114905, Dragonfly
44, the Bullet Cluster and Abell 520 (though the Abell 520 is
still under investigation and debated), provide strong
confirmation of the DM's existence as a component of galaxies, but
the latter two obscure the nature of DM on whether it is
collisional or collisionless. Furthermore, the existence of both
DM dominated and DM deficient (NGC 1052-DF2) galaxies indicates
that DM can be present or absent, contradicting MOND on the
universal modification that would affect every object equally.
Even if MOND succeeds in explaining the galactic rotation curves,
faces a big challenge when confronted with a broader range of
phenomena, at cluster and cosmological scales. It lacks of a
viable and GW170817 compatible relativistic framework, it cannot
explain the damping of the BAO wiggles, misinterprets the CMB
peaks (relation between the second and third), overproduces early
structures and most importantly requires invisible matter in
galactic clusters. Therefore MOND attempts to fit cosmological
observations, but still requires unseen matter, something that
undermines its own core idea. Additionally, the spin problem, that
initially challenged $\Lambda$CDM as well, has been successfully
resolved through spin segregation, after we included feedback
processes like supernovae feedback and active galactic nuclei
feedback. MOND does not provide us with any analogous mechanism,
on the contrary, it aggravates the problem. Future observations
from DESI, Euclid and Rubin Observatory, will continue to examine
$\Lambda$CDM's predictions, with more precision. MOND was produced
to interpret galaxy rotation curves, and only succeeds with fine
tuning. $\Lambda$CDM succeeds in explaining almost everything
across the entire cosmic landscape, save some galactic scale
problems like the cusp core problem. Having the latter problems in
mind, combined with the obscure nature of DM from the Abell 520
and Bullet cluster, in this work we also presented a new viewpoint
of MOND theories. MOND theories should not be viewed as a solution
to galactic scale physics, but they should be viewed as a
phenomenological description of DM at galactic scales.
Specifically, DM should reproduce the MOND behavior at radii near
the core and far from the core of galaxies. One framework capable
of achieving this is scale-dependent self-interacting
multi-component DM, which can exhibit both collisional and
collisionless behaviors, depending on the scale. Such examples of
DM exist in the literature in the form of mirror DM
\cite{Kobzarev:1966qya,Hodges:1993yb,Foot:2004pa,Berezhiani:2003wj,Silagadze:2008fa,Foot:2000tp,Chacko:2005pe,Berezhiani:2000gw,Blinnikov:2009nn,Mohapatra:2001sx,
Blinnikov:1982eh,Blinnikov:1983gh,Foot:2016wvj,Foot:2014osa,Foot:2014uba,
Foot:2004pq,Foot:2001ft,Foot:2004dh,Foot:1999hm,Foot:2001pv,Foot:2001ne,Foot:2000iu,Pavsic:1974rq,Foot:1993yp,Ignatiev:2000yy,Ignatiev:2003js,
Ciarcelluti:2004ik,Ciarcelluti:2004ip,Ciarcelluti:2010zz,Dvali:2009fw,Foot:2013msa,Foot:2013vna,Cui:2011wk,Foot:2015mqa,Foot:2014mia,Cline:2013zca,Ibe:2019ena,
Foot:2018qpw,Howe:2021neq,Cyr-Racine:2021oal,Armstrong:2023cis,Ritter:2024sqv,Mohapatra:1996yy,Mohapatra:2000qx,Goldman:2013qla,Berezhiani:1995am,Oikonomou:2024geq},
and the galactic physics of such scale-dependent DM profiles was
analyzed in Refs. \cite{Oikonomou:2026vkp,Oikonomou:2025bsi}. Thus
MOND should not be viewed as the solution to the missing matter
problem, but it should be viewed as the phenomenological behavior
of DM at galactic scales. This puts a different viewpoint for a
rather failed phenomenological framework like MOND.

\end{document}